# Uncovering contributing factors to interruptions in the power grid: An Arctic case

Odin Foldvik Eikeland [a], Filippo Maria Bianchi [b,c], Inga Setså Holmstrand [d], Sigurd Bakkejord [d], Sergio Santos [a], and Matteo Chiesa [a,e,*]

[a] UiT the Arctic University of Norway, Department of Physics and Technology, 9037 Tromsø, Norway.
[b] UiT the Arctic University of Norway, Department of Mathematics and Statistics, 9037 Tromsø, Norway.
[c] NORCE, the Norwegian Research Centre, 9037 Tromsø, Norway.
[d] Arva Power Company, 9024 Tromsø, Norway.
[e] Laboratory for Energy and NanoScience (LENS), Khalifa University of Science and Technology, Masdar Institute Campus, 127788 Abu Dhabi, United Arab Emirates
[*] Corresponding Author: matteo.chiesa@uit.no (M. Chiesa)

## Abstract

Electric failures are a problem for customers and grid operators. Identifying causes and localizing the source of failures in the grid is critical. Here, we focus on a specific power grid in the Arctic region of North Norway. First, we collect data pertaining to the grid topology, the topography of the area, the historical meteorological data, and the historical energy consumption/production data. Then, we exploit statistical and machine-learning techniques to predict the occurrence of failures.

We interpret the variables that mostly explain the classification results to be the main driving factors of power interruption. We are able to predict 57% (F1-score 0.53) of all failures reported over a period of 1 year with a weighted support-vector machine model. Wind speed and local industry activity are found to be the main controlling parameters where the location of exposed power lines is a likely trigger. In summary, we discuss causing factors for failures in the power grid and enable the distribution system operators to implement strategies to prevent and mitigate incoming failures.

# 1 Introduction

Electric failures affect everyone connected to the power grid, from single homeowners to large industries [1], [2], [3], [4]. The distribution system operator (DSO) is contractually obliged to robustly supply electricity and compensate customers affected by any possible power interruptions [5]. Nevertheless, DSOs need the proper energy management plans, technology, and infrastructure to meet the demand. Electric failures alone can turn into serious losses whose estimates range from $22 to $135 billons annually [6], [7], [8]. Moreover, failures might have complex and adverse socio-economic consequences in communities heavily reliant on the electricity supply to satisfy their basic needs [9], [10].

Here we focus on a specific grid in the Arctic region of North Norway. In this context, the expansion of local industries connected to the power grid is driven by the processing and exporting of fish to the worldwide market. An important requirement is that these production lines operate at full capacity during the high season. To this end, several automated components have been introduced that are critically dependent on reliable power quality. Even minor electrical disturbances lasting only a few seconds could stop the entire production line for a significant amount of time. In particular, for every short-term power interruption that occurs, 40 minutes to 1 hour might pass before resuming production.

In most communities, multiple sources could provide backup electricity supply if one line is interrupted [11]. However, the power grid considered in our study has a radial distribution system where customers connected to the grid do not have any sources to provide backup power in case of a Failure. It is therefore fundamentally important to develop strategies to increase the reliability of the power grid that ensure the growth of local industries.

One way to deal with this situation is to build a new power grid and design it to handle the current power requirements. However, this is costly, time-consuming, has a huge environmental impact, and contradicts the vision of better utilizing the current electricity grid [12], [13]. Moreover, if the current companies ever stop or reduce their activities in the future, the power companies will get an over-dimensioned distribution network. An alternative solution is to limit the failures and strengthen only the most vulnerable parts of the grid but this implies first identifying the factors triggering power interruptions.

The identification of such factors, and their location in the grid, has proven to be a major challenge for the DSO [14]. The increased availability of energy-related data by the DSOs however, makes it now possible to exploit data science techniques to develop strategies to improve the reliability of the power grid [15]. Some effort has been put in this direction where the performance of the proposed solutions has been applied on benchmark power systems [16], [17]. Results indicate that extreme weather conditions are often the major cause of failures in the power grid. In fact, some have recently employed an approach to investigate the relationships between weather variables and power failures with the use of machine learning (ML) techniques [6]. However, it is likely that other than weather conditions, human activity also plays a key role. In this work, we explore a wider spectrum of explanatory variables and propose that, rather than assuming that harsh weather conditions are the only source of failures, both the topography of the energy grid and the time series of energy flows are considered. We exploit statistical and ML techniques to detect the causes of power interruptions by including explanatory variables that are divided into two groups: meteorological and energy-related (consumption-and production) data.

Our aim is to discover the factors that trigger failures. In close collaboration with the local DSO, we investigate a real-world power grid that has problems with the reliability of the power supply.

The work is structured as follows. In section 2 we review some of the most relevant work on the topic. In section 3 the specific power grid analyzed here is presented together with the report of failures from the local DSO. In Section 4 we present the methodology for deriving the dataset used in the statistical analysis. Here the features are analyzed and the different ML techniques used for classification and failure detection are presented. In section 5 we investigate the quality of the models in relation to the identification of the cause of failures. Conclusions are given in section 6.

# 2 Related works

## 2.1 Detecting failures caused by weather conditions

While prior studies on this topic are limited, harsh weather conditions are believed to be an important source of failures, and several studies have been conducted to address the impact of weather events on power quality.

In the study by Owerko et al. power failures in New York City were predicted based on weather conditions [6]. The authors deployed a Graph Neural Network (GNN) to adequately process weather measurements close to the power grid in order to determine the likelihood of power failure. Results showed a 1.04% error in the prediction.

The impacts of weather variations and extreme weather events on the resilience of energy systems were investigated by the authors in [18]. The authors developed a stochastic-robust optimization method to consider both low impact variations and extreme events. The method was applied on 30 cities in Sweden. One finding indicates a 16% drop in power supply reliability due to extreme weather events.

The authors in [17], provide a framework for quantifying and modeling the resilience of power systems that address the challenge of high wind incidences. The proposed algorithm involves a matrix-based approach to account for all possible routes from generators to loads. This is done by considering the load importance before and after a disruption incident. The authors illustrate the effectiveness of the proposed approach on the IEEE 14-Bus system.

The authors in [16], address the challenges posed on the energy grid by frequent extreme weather events. The authors propose an online spatial risk analysis that could provide an indication of the evolving risk of systems placed in regions subject to extreme events. The authors used a Severity Risk Index with the support of real-time monitoring. The index considers the spatial and temporal evolution of the extreme event, system operating conditions, and the degraded system performance during the event. The online risk analysis was embedded in a sequential Monte Carlo simulation to capture the variability of the spatiotemporal effects of extreme weather events.

[19] is a review of existing methodologies where they assess the impact of weather on power system resilience. Critical infrastructure and plans for improving resilience in cases of extreme weather events are also provided.

The authors in [20] together with the Italian Transmission System Operator (TSO), proposed an approach to improve grid resilience in natural disastrous events. The approach deploys flexibility tools that combine the flexibility margins of the networks to those of generation and load. The authors propose a flexibility tool to improve the resilience of the power system in case of severe weather conditions.

All the studies mentioned so far consider severe weather events only and disregard other factors such as heavy energy load caused by human-related activities. Additionally, most test their methodologies on benchmark datasets rather than on real case-studies.

## 2.2. Alternative approaches for failure detection

A methodology to predict power failures by analyzing advanced measurement equipment such as Power Quality Analyzers (PQAs) and Phasor Measurement Units (PMUs.) has been proposed [21]. Here, the measurement equipment has high resolution and generates a vast amount of data. The study used real-world measurements from nine PQA nodes in the Norwegian grid to predict incipient interruptions, voltage dips, and earth failures. The authors find incipient interruptions easiest to predict while reporting that earth failures and voltage dips are more challenging to predict.

The challenge of detecting earth-failures, a challenge for the DSOs for decades, has been recently addressed [22]. In principle, their findings show that by integrating advanced metering infrastructure with a distribution management system, failures can be detected. However, the proposed solution could only be used if the DSOs could access-and knew how to use the OpenDSS software.

The challenge of the increasing strain on the power grid with increasing electrification has also been investigated [23]. Here the challenge with the implementation of renewable energy technologies with intermittent power production is also discussed. The authors proposed possible monitoring solutions and failure-predicting methodologies by using statistics and field-measurements from the Norwegian power grid. While ML approaches were identified as candidates, the authors did not propose any. No time resolution and/or duration was considered either in terms of developing a robust failure detection and prediction methodology.

# 3. Case study

## 3.1 The case of SVAN22LY1

The power grid is a radial distribution system where the power is distributed from the south (69.257°N, 17.589°E) towards the north. Arva Power Company, the DSO of the power grid, has named this specific radial as SVAN22LY1. See Figure 1 for a map showing an overview of the whole SVAN22LY1 grid, indicated by green dots.

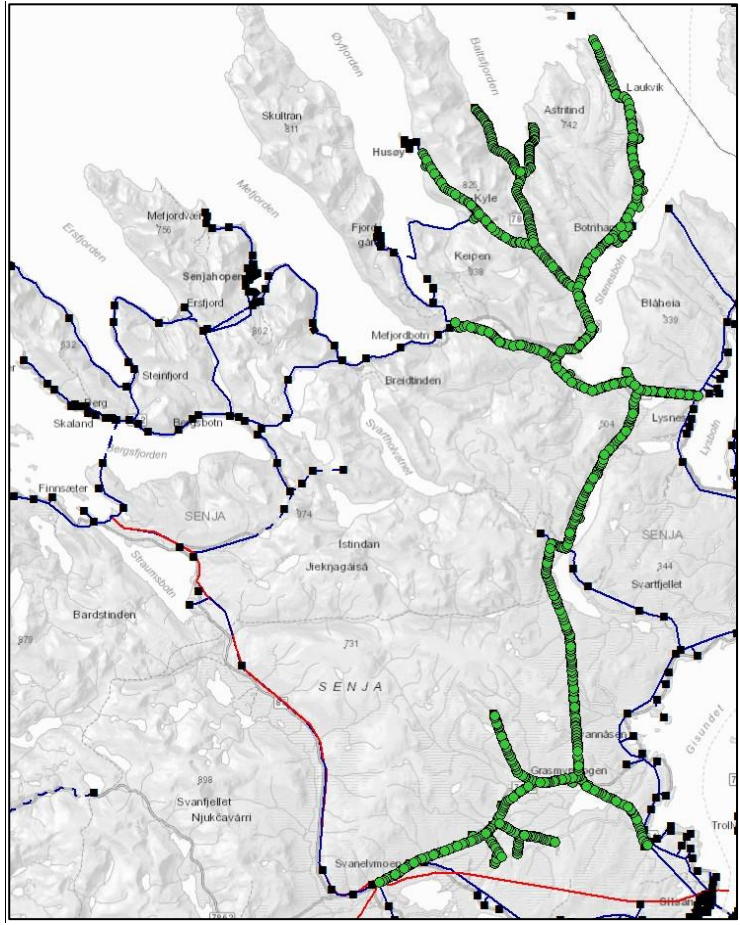

*Figure 1: The SVAN22LY1 power grid. The power is distributed towards the north from the south. Each green dot represents a unique position of a utility pole.*

The SVAN22LY1 grid spans over 60 kilometers from south to the northernmost point and has several branches to various communities towards the north. There are 978 unique utility poles (marked by green dots in Figure 1) that support the power lines. The black boxes in Figure 1 represent the electric transformer stations connected to the power grids. The red lines represent a power grid with an operating voltage of 66 kV, while the blue lines represent a power grid with an operating voltage of 22 kV. The SVAN22LY1 radial grid covered by green dots has 22 kV operating voltage [14].

The largest customers connected to the SVAN22LY1 grid are located at the end of the northernmost point of the radial (69.546°N, 17.657°E). The total energy demand in the particular community is characterized by a combination of load profiles from two sectors: households and industry. The particular feature of the rural community is that industry accounts for more than 50% of the total energy consumption [24]. The industry has electrical machines that are sensitive to stable power quality, and minor power interruptions could bring the production line to a halt.

The SVAN22LY1 power grid is in an area that is characterized by typical Arctic conditions with cold, long winters and is heavily exposed to harsh weather conditions throughout the whole year. The harsh weather conditions can result in voltage drops and power interruption, in addition to the increased power demand from end customers. For example, strong wind gusts can make the power lines between the utility poles collide. When phases of the power lines collide, it will cause interruptions in the power supply [25].

## 3.2 Reported Failures in the SVAN22LY1 power grid

Every failure in the power system is logged as an incident in the Norwegian reporting system FASIT (Failure-and interruption statistics within the national power grid) [26]. Each DSO company in Norway is required to have a FASIT reporting system. The purpose of FASIT is to provide information about the delivery reliability of the Norwegian power system. The report provides both information about historical delivery reliability and information for estimating future expected delivery reliability. To be able to obtain information on the delivery reliability, the report should contain information on:

- Operational disturbances
- Planned disconnections that have led to power interruptions (both planned notified disconnection and planned unannounced disconnection)

The operational disturbances should cover information about each failure, on which grid the failure occurs, and if known, the reason for the failure. The operational disturbances should also cover an end-user focus by reporting the duration of the interrupted power. In addition, the amount of not-delivered power and the compensation for not-delivered power are reported [26].

The FASIT system also holds information about short-term power interruptions that typically are less than one minute. Although the power interruptions are very short, they have major consequences for the local industries, as these events may temporarily halt the entire industry – resulting in economic losses.

In this study, the FASIT report for the SVAN22LY1 grid is provided by the Arva Power Company [14]. In 2020 (January to the end of November), the FASIT system reported 54 incidents, distributed over 44 different hours (several failures can occur within one hour). Figure 2 shows the reported failures in 2020. The distribution of failures during the year is not uniform, and most failures were reported in September month followed by January and July. In June, no failures were reported. In this study, the period from the first of January 2020 to the end of November 2020 is investigated.

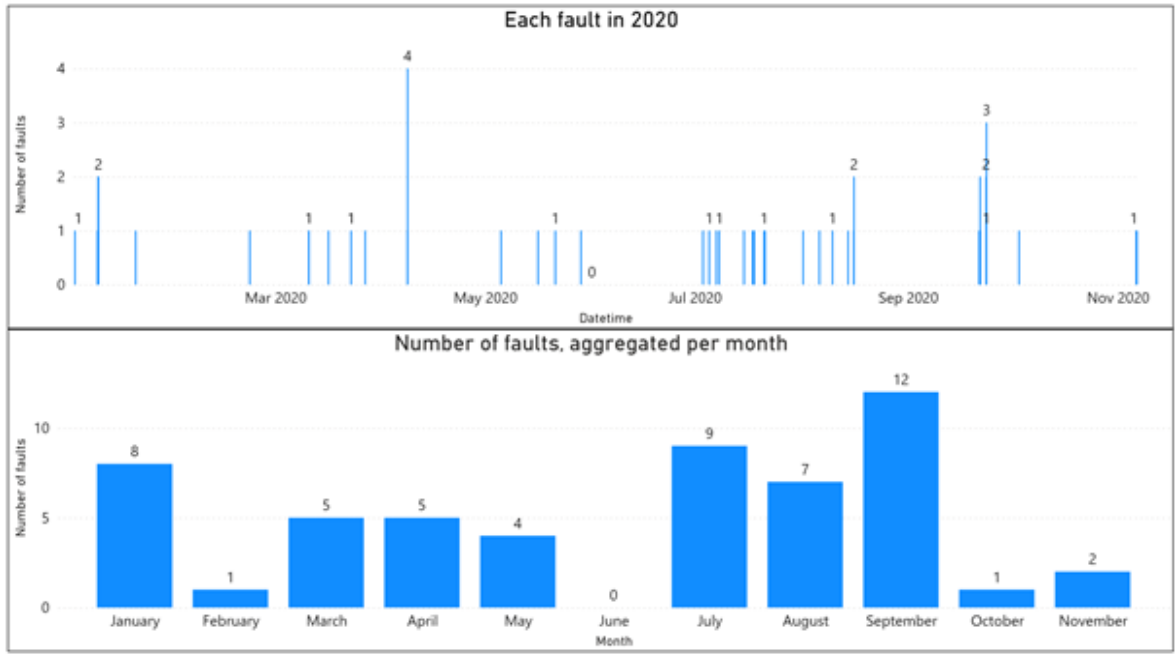

*Figure 2: Distribution of Failures in 2020 in the SVAN22LY1 power grid, logged in the FASIT report of Troms Kraft*

The short-term failures are reported in the FASIT system, but it has proved to be a major challenge to identify what caused the error, and the location in the grid of what triggered the failure event. In this study, the location of what triggered the failure is assumed based on a set of parameters and is not fully localized. This is a potential future work, and this study prepares the ground for such a follow-up study.

If the location and the causes for each short-term failure are detected, the DSO can take measures to avoid power failure-incidents in the future.

# 4  Method

## 4.1 Construction of the dataset

To detect the potential contributing factors for failures, all potential variables of interest are collected. Then, statistical methods are used to select the relevant variables that explain the failures in the power grid. The possible variables that could cause failures in SVAN22LY1 are divided into two groups: Weather-related failures and energy-related (consumption-and production) failures.

### 4.1.1  Localizing weather-exposed areas in SVAN22LY1

As the SVAN22LY1 grid spans over a large area, the weather conditions can change significantly depending on geographic location. To detect the regions in the power grid that are more likely exposed to weather-related failures, some assumptions are required. One is to assume that higher elevation increases the probability of being exposed to harsh weather conditions, such as strong wind. Indeed, utility poles at high altitudes are often in mountainous areas where there is no vegetation that can protect from the wind.

The altitude map for the area of interest is obtained from the Norwegian Mapping Authority [27]. The altitude map is a digital surface model (DSM) with a spatial resolution of 1-meter [27], [28]. To obtain the altitude for the different utility poles, the geographic coordinates for each pole were mapped into the DSM. The poles at the lowest altitude are at sea level (0 MASL (Meters Above Sea Level)), whereas the utility pole at the largest elevation is at 485 MASL.

We group the altitudes into 5 categories: 1 (0-100 MASL), 2 (100-200 MASL), 3(200-300 MASL), 4(300-400 MASL), and 5(400-485 MASL). We assume that the regions more exposed to harsh weather, are category 4 and 5 (300 to 485 MASL).  The utility poles in high altitude regions are highlighted in Figure 3.

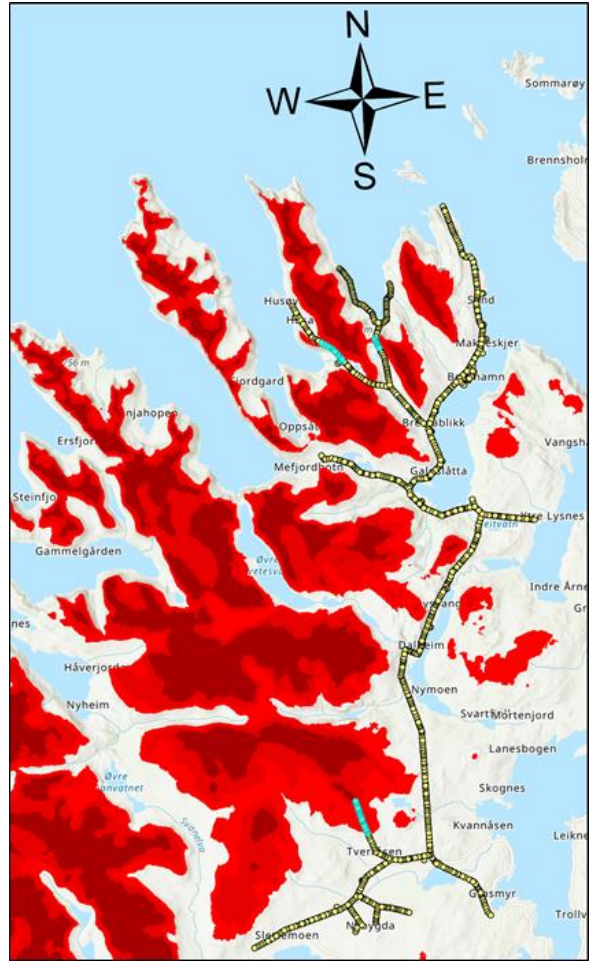

*Figure 3: DSM altitude map of the area of interest. The weather-exposed areas in the power grid are highlighted in blue.*

The second piece of information that is useful to consider is the distance between the utility poles. We assume that the longer the length between two poles, the higher is the likelihood that the cables will collide under harsh weather conditions, resulting in power interruptions. The distance between adjacent utility poles is found by calculating the geodesic distance between the Lat-Long coordinates of each utility pole.

Based on the two criteria defined above (altitude and long distances between poles), we selected 3 regions that are particularly at risk. There is no guarantee that these are the areas where the weather-related failures have been triggered. However, by identifying these areas we can significantly narrow down the candidate regions that triggered the failures.

The resulting exposed areas (EAs) are:  EA1, EA2, and EA3. Information about each region is given in Table 1.

*Table 1: Properties of selected weather-exposed areas in SVAN22LY1 power radial*

|  | EA1 | EA2 | EA3 |
|---|---|---|---|
| **Coordinate (deg.)** | 69.523N, 17.779E | 69.520N, 17.726E | 69.321N, 17.729E |
| **Elevation (MASL)** | 300 | 360 | 485 |
| **Longest distance between poles (m)** | 122 | 136 | 121 |

### 4.1.2  Exposed areas and weather-variables from AROME-Arctic

The selected EAs are located kilometers apart (see Figure 3), and there is no weather station located in the proximity of either region. Therefore, to collect relevant weather-data in the specific regions, we use the AROME-Arctic weather model. This model is developed by the metrological institute of Norway (MET). The model is a reanalysis model that has run since November 2015 and has a spatial resolution of 2.5 kilometers. MET has developed this dedicated model due to several challenges to weather conditions that are unique in Arctic regions, such as polar lows and icing [29]. The model covers the latitudes from 66N to 88N and longitudes from -18E to 80E.

To import the relevant weather variables, we use the coordinates from the EAs as input to the AROME-Arctic and collect the following variables: *Wind direction, wind strength, precipitation, relative humidity, temperature, and pressure*. The AROME data are in hourly resolution and gives 8002 samples from 01.01.2020 to 30.11.2020. One way is to model the weather variables for each EA, giving totally 18 weather variables (6 for each location) that must be investigated. However, the weather variables are very similar in each EAs, and therefore we choose to average the variables from all EAs, resulting in one dataset of 6 weather-variables that are analyzed.

### 4.1.3  Detecting non-weather-related failures

In our analysis, we want to consider the possibility that some failures are not caused by weather phenomena. To model these effects, the power consumption data from the largest industry connected to SVAN22LY1 are collected. Two variables are included: The average consumption over one hour and the difference in minimum and maximum power during this hour. The minimum/maximum power is logged every 30. Second. The energy data available from the DSO have different temporal resolutions, and to perform analyzes, all variables must be transformed to the same resolution. The meteorological data from AROME-Arctic are only available in hourly resolution, and therefore the dataset consisting of all variables transformed into hourly resolution. To transform the minimum/maximum power data which are logged every 30. Second into hourly resolution, the largest difference in minimum/maximum power during the particular hour is selected. A large difference in minimum/maximum power consumption corresponds to a high level of activity at the local industries.

In addition, the energy consumption data from one electrical transformer in the southern part of the SVAN22LY1 are included (referred to as "Transformer south" in the following).  This electrical transformer station provides information about the energy consumption for all customers connected to it, and the customers that are not directly connected to the SVAN22LY1 grid. Therefore, this electrical transformation station could provide insight into whether other types of energy consumption patterns than the local industry could affect the power quality in SVAN22LY1. The energy consumption data from Transformer south are in hourly resolution.

There is a hydropower station connected directly to the SVAN22LY grid, and it would be of interest to investigate whether local hydropower production could affect the power quality. For instance, when there is low local production, electricity must be imported from other supply-sources farther away from the power grid. This will result in transmission losses. Transmission losses refer to the loss of energy when it is transported by the power line. The longer is the distance of a power line, the larger is the transmission losses. Therefore, if the hydropower station does not produce local energy, and the energy must be transported over a long distance, the transmission losses will result in a reduced amount of energy to the customers. Consequently, this could have a negative impact on the local power quality. In addition, if electricity must be imported from other sources, neighboring parts of the power grid could be overloaded and in the worst-case result in failure [30].

We also include variables that could affect the power quality, even if they are not coming from measurements collected in SVAN22LY1. In particular, we include the consumption data from an external industry, which is located 30 kilometers away from the electrical transformer located in the south (Transformer south). This industry is the largest customer in the entire power grid of the DSO. The particular industry has a consumption profile that puts a heavy strain on the grid and could affect the power quality in a large area [14].

Three measurements are collected from this external industry: Reactive power and energy consumption. The reactive power is measured at two different electric transformer stations, T1 and T2.

The location of the source of all variables used in our analysis is visualized in Figure 4.

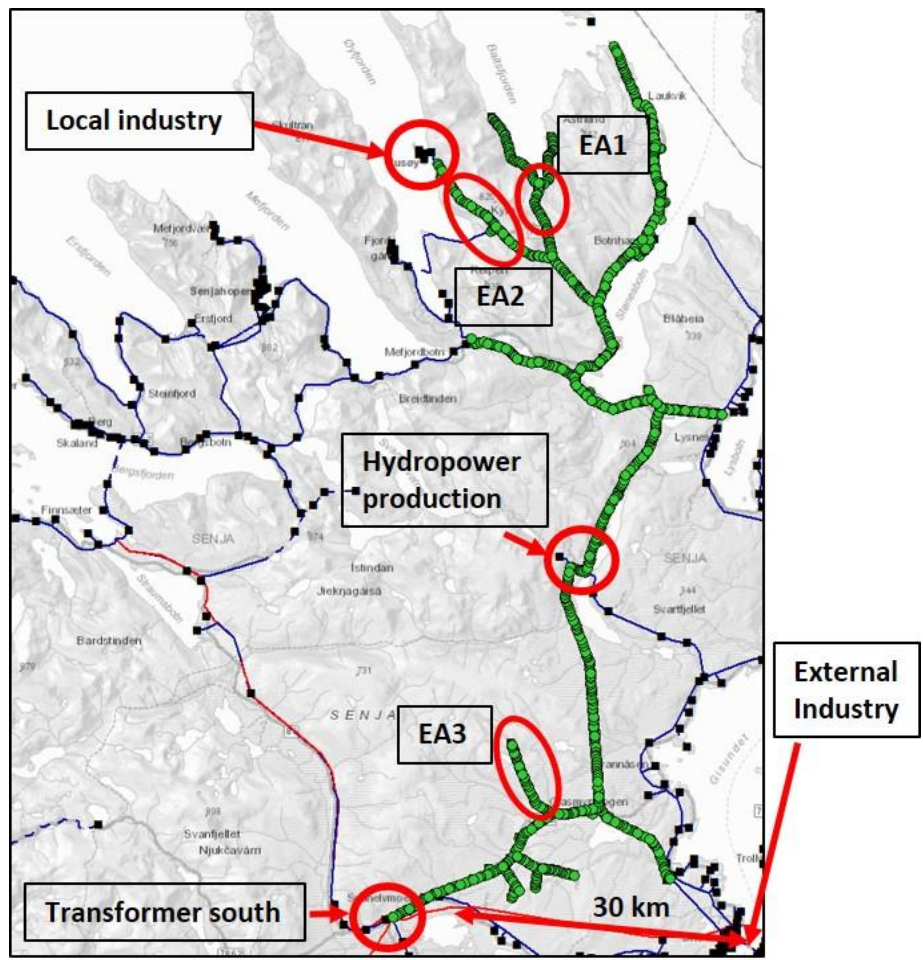

*Figure 4: The SVAN22LY1 grid with locations for all the variables included for statistical analyzes. The Figure shows the SVAN22LY power grid marked by the green dots that represent the positions of the utility poles. Each EAs are highlighted. EA1 and EA2 are in the northern part of the grid, and EA3 is in the southern part. The electrical transformer station in the south is located approximately 30 kilometers away from the external industry. The hydropower station is in the middle of the power grid and supports the communities with electricity. The largest industry connected to SVAN22LY1 is in the northernmost part of the grid.*

The final dataset has 8002 samples with hourly resolution data from 01.01.2020 to 30.11.2020. The variables analyzed are given in Table 2:

*Table 2: Variables analyzed to detect Failures in the SVAN22LY1 power grid.*

| Feature | Weather-variables |
|---------|-------------------|
| 1 | Wind direction |
| 2 | Wind speed of gust |
| 3 | Temperature |
| 4 | Air pressure |
| 5 | Relative humidity |
| 6 | Precipitations |
| **Feature** | **Non-weather variables** |
| 7 | Local industry: Energy consumption |
| 8 | Local industry: Difference minimum/maximum power consumption |
| 9 | Hydropower production |
| 10 | External industry: Reactive power (T1) |
| 11 | External industry: Reactive power (T2) |
| 12 | External industry: Energy consumption |
| 13 | Transformer south |

## 4.2 Properties of feature classes

Once the dataset with all variables is constructed, we model each data point $y$ as a binary variable where $y=1$ indicates that a power failure occurred, and $y=0$ indicates that no failure happens (i.e., the power grid operates as it should). The binary variable divides the dataset into two classes: A minority class (totally 44 samples with failures) and a majority class (totally 7958 samples with non-failures, i.e., normal conditions).

To visualize the distribution of the values in each variable across the different classes, we use a Kernel Density Estimation (KDE) [31]. We used a Gaussian kernel in the KDE to estimate the distributions.

The normalized Gaussian KDE functions for each feature are provided in Figure 5. If the estimated density functions show significant differences between the failure class (red) and the normality class (blue), this indicates that a specific feature behaves differently in the two classes and can be used to discriminate among them. Looking at the distributions might help to assess which features are more important to identify the failures in the power grid. On the other hand, less-important features have a similar distribution in both classes and are arguably less useful to discriminate among them.

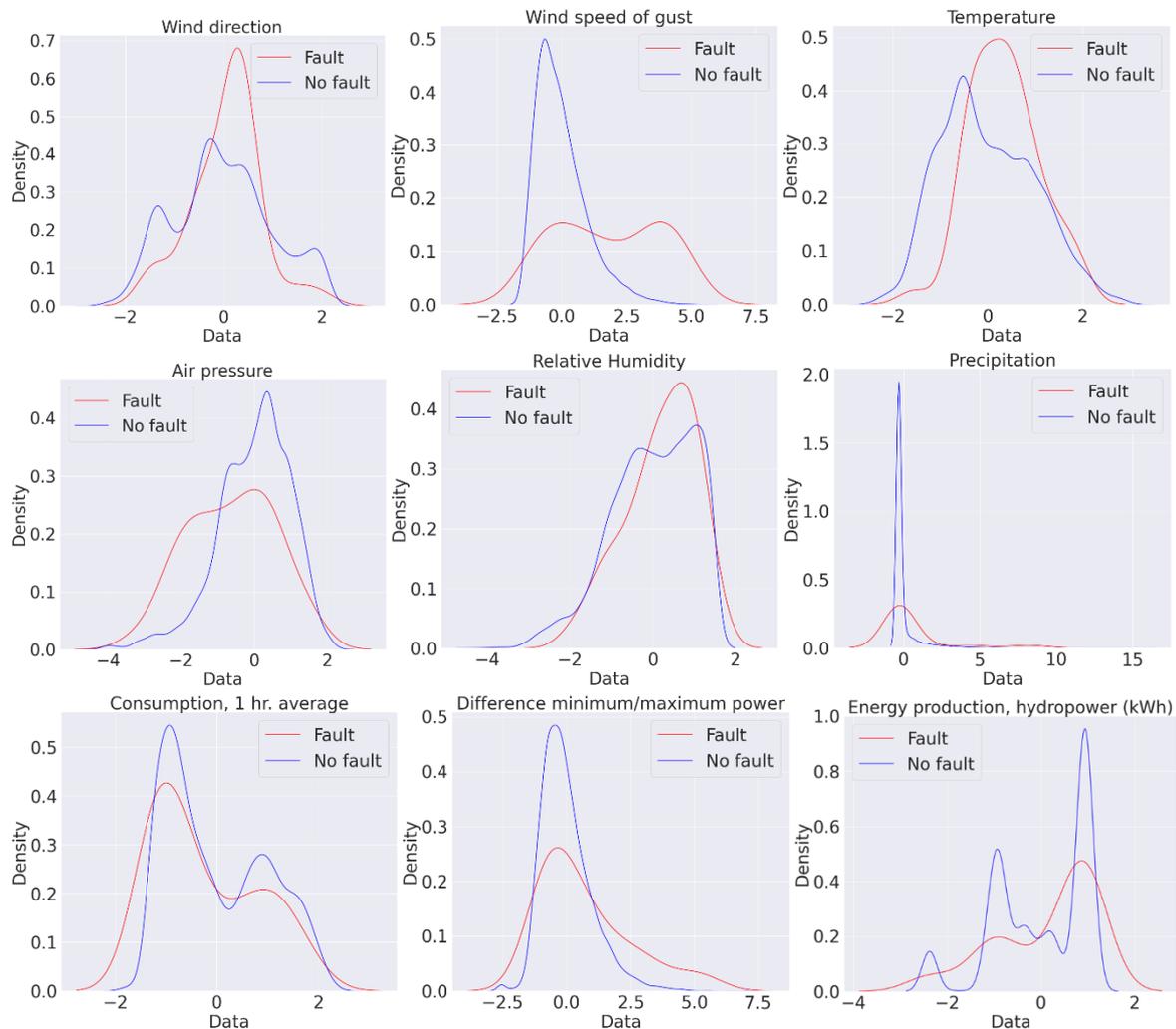

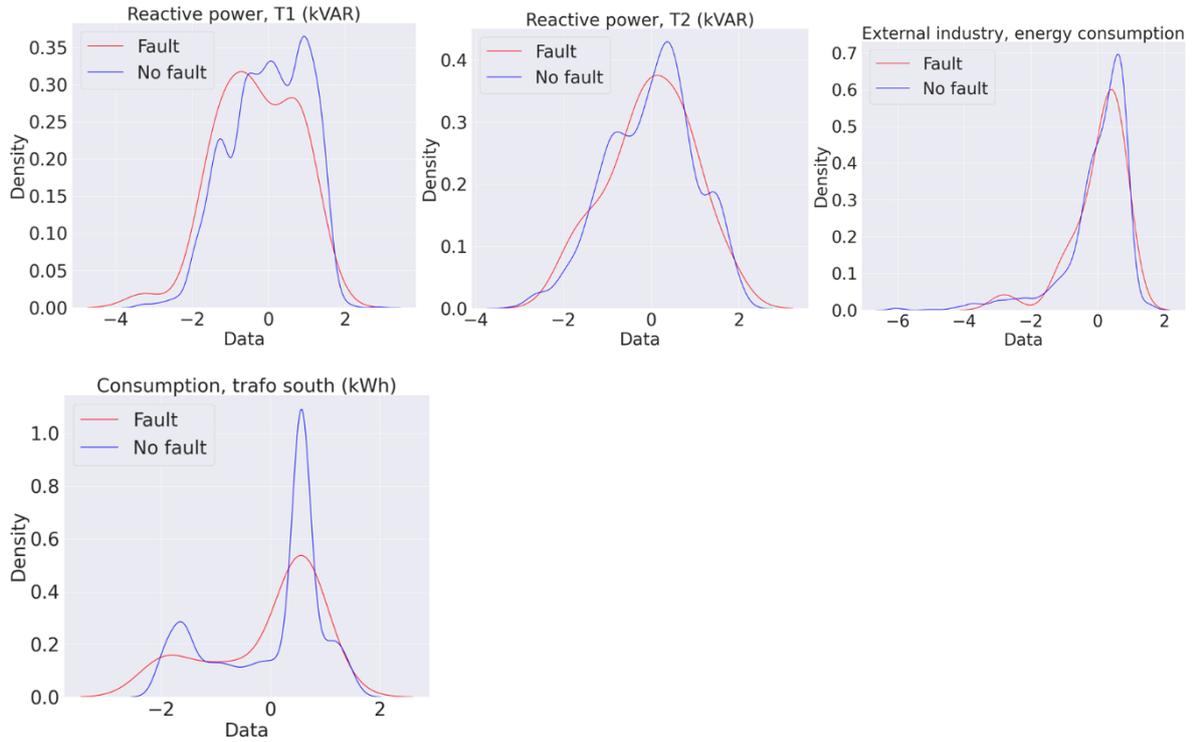

*Figure 5: Normalized Gaussian Kernel Density Estimate of minority class (red) and majority class (blue) for each feature.*

It is clear from the density plots that it is very difficult to discriminate the minority class from the majority class based on these features, as the distributions are similar. From the density plots, it seems that wind speed is the most discriminating feature since the distributions are different in the two classes. However, it is difficult to verify that wind is the only feature causing failures by investigating the KDE distributions.

To quantify the separation between the distributions numerically, the Kullback-Leibler (KL) divergence is computed for each variable. If the KL divergence is higher-the larger is the discrepancy between the two distributions [32]. The value of the KL divergence is given in Table 3.

*Table 3: KL divergence for each variable*

| Variable | KL divergence |
|---|---|
| Wind dir. | 2.28 |
| Wind speed | **3.87** |
| Temp. | 2.21 |
| Air pres. | 1.61 |
| Rel. Hum. | 2.35 |
| Precip. | 0.82 |
| Local industry: Energy consumption | 3.83 |
| Local industry: Diff. min/max power. | 2.03 |
| Energy production, hydropower | 1.62 |
| External industry: Reactive Power (T1) | 3.58 |
| External industry: Reactive Power (T2) | 1.67 |
| External Industry: Energy consumption | 2.69 |
| Consumption, Transformer south | 3.83 |

The KL-divergences confirms that wind speed is the most discriminating feature, as indicated from the distribution plots in Figure 5. However, since the values are similar, it is difficult to verify that wind is the only feature causing failures in the SVAN22LY1 grid.

It is believed that rapid changes (big changes within seconds/minutes) in power consumption at large industries could affect the power quality in a wide area [14]. The rapid changes will last for only seconds/minutes, and the hourly resolution data will not cover such events. Therefore, it will be of interest to investigate the potential impact of such rapid changes by analyzing higher-resolution consumption data.

Some higher-resolution consumption data for the external industry are available and are provided from the DSO. The resolution is 1 minute, resulting in a dataset of 480,120 samples. The detailed analysis of the high-resolution data is given in the supplementary chapter. The supplementary is available upon request by the author.

### 4.2.1 Visualization of majority class and minority class

To get an overview of how the majority class (normal conditions) and minority class (failures) are distributed, the features are visualized as a scatter plot. Figure 6 provides examples of how the different features are spread for each class.

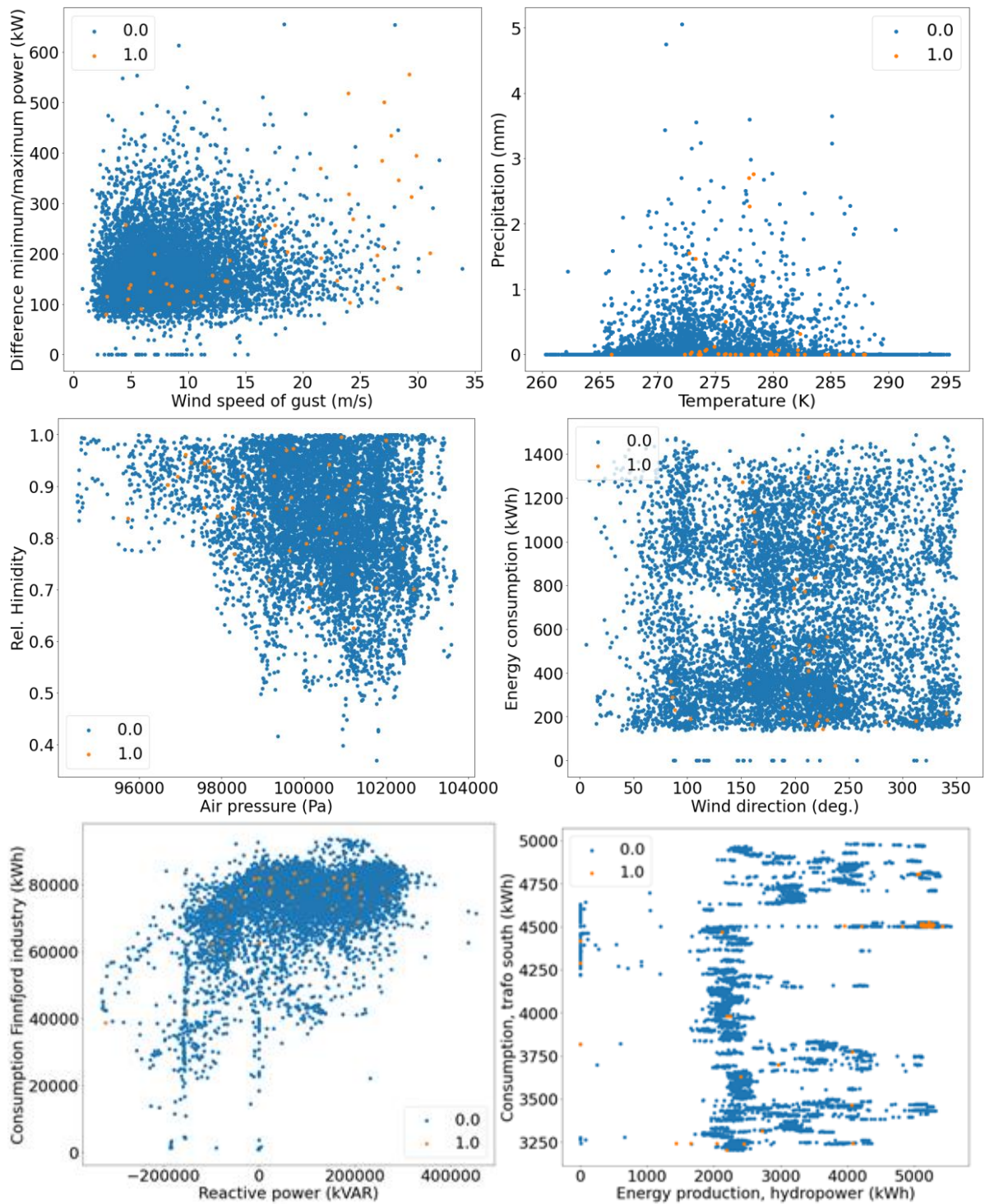

*Figure 6: Scatterplot of the two classes for different features. One majority class of non-failures, and one minority class with failures (indicated by 1).*

Figure 6 shows two major challenges regarding distinguishing the minority and majority classes. First, the classes are very imbalanced. There are only 44 incidents with failures, but 7957 non-failures, i.e., normal conditions. Secondly, there is no clear separation between the two classes.

The minority class (failures) are mixed with the majority class (normal conditions) repeatedly. However, the scatterplot of *wind speed of gust* and *difference in maximum/minimum power* (upper left), shows some anomalies in the minority class (to the right in the scatterplot). This indicates that the SVAN22LY1 power grid might experience more failures when there is a high wind speed of gust (above 25 m/s). This is also clear from the second density plot in Figure 5, and the KL-divergence in Table 3.

### 4.3 Classification and outlier detection strategy

In this study, the failures in the SVAN22LY1 grid are identified using different machine-learning (ML) models. We frame the failure identification as two different tasks: classification and anomaly/outlier detection.
The ML model used for classifications is the Support Vector Machine (SVM). To perform anomaly detection, we use four types of one-class classification models (One Class SVM, Isolation Forest, Elliptic Envelope, local outlier factor).

The SVM is a popular ML model used for classification problems [33], [34]. The objective of the linear SVMs is to find the optimum hyperplane that separates two classes with a maximum margin [35]. A non-linear SVM finds the separation boundary in kernel space. As the relationship between the minority and the majority class are very mixed, the original SVM model might not be able to distinguish between the two classes. When the two classes are too imbalanced, the SVMs can choose to sacrifice all minority class samples and achieve a good classification for the majority class instances. Consequently, the SVM assumes that everything is non-failures and is not able to predict any failures that occur. One way to tackle the challenge is to weigh more the classification errors on the minority class [36].
This gives a weighted SVM that is weighting the False Positive and False Negative errors differently. To investigate the properties of the weighted SVM to tackle the imbalance, the False Negative errors are weighted more than the False Positive errors.

The one-class models are commonly used for classification tasks with an imbalanced class distribution and have shown to be effective when there are very few examples of the minority class [37]. The dataset analyzed in this study has very few examples of the minority class (44 out of 8002 samples) and the one-class algorithms could be suitable.

Another technique to deal with class imbalance is to balance the original dataset. We apply Synthetic Minority Oversampling Technique (SMOTE) [38]. The SVM is thereafter trained on the synthetic dataset before classifying the original-and imbalanced dataset.

A more detailed explanation of the models used (one-class algorithms and SVM) and the SMOTE approach is deferred to the supplementary material.

### 4.3.1 Training and testing

This section explains how to train the ML models and use the trained model on the test set. We split the data into 85% training and 15% test. This gives a training set of 6801 samples and a test set of 1200 samples. The training set is used to fit the model parameters by minimizing the prediction loss. Once the optimal model is found, its performance is evaluated on the test set. We pre-process the data by normalizing the samples. The normalization is done by subtracting the mean and dividing by the standard deviation computed on the training set.

### 4.3.2 Measuring classification performance

As the dataset is very imbalanced, it is difficult to obtain high accuracies from the different ML models. Therefore, to evaluate the performance of the classification, the F1 score is used to measure the performance on the test set (1200 samples). The F1 score is interpreted as the weighted average of the precision and recall, where the highest score is 1 (able to predict all incidents), and the lowest score is 0 (not able to predict any incidents) [31]. The F1 score is defined as

$$F1 = 2 * \frac{\text{precision} * \text{recall}}{\text{precision} + \text{recall}} = \frac{\text{TP}}{\text{TP} + \frac{1}{2}(\text{FP} + \text{FN})}, \tag{1}$$

where TP=number of true positives, FP=number of false positives, and FN=number of false negatives. In this study, a positive is a Failure and a negative is non-Failures (i.e., normal conditions). The F1 scores for all models are provided in Table 4 in the result section.

In addition, the sensitivity score (proportion of positives that are correctly identified) and specificity score (proportion of negatives that are correctly identified) are reported. The sensitivity and specificity scores are defined as:

$$\text{TPR} = \frac{\text{TP}}{\text{P}} \tag{2}$$

$$\text{TNR} = \frac{\text{TN}}{\text{N}} \tag{3}$$

Where P, N is the total number of positives and negatives, respectively.

# 5 Results

## 5.1 Classification scores

The test set consists of 1200 samples divided into 7 failures and 1193 non-failures. The resulting F1 scores for all models are provided in Table 4.

*Table 4: Classification results for each model*

| Model | F1 score |
|-------|----------|
| One-class SVM | 0.101 |
| Isolation forest | 0.113 |
| Elliptic envelope | 0.169 |
| Local outlier factor | 0.138 |
| SVM | 0.000 |
| Weighted SVM | 0.533 |
| SMOTE-SVM | 0.444 |

The results show that the one-class models achieve low classification performance, as indicated by the F1 score. For this reason, the one-class models are not further considered in this study. The F1 score of the original SVM is 0, which indicates that the original SVM assigns each data point to the majority class. When there is a strong imbalance between the two classes, the original SVM finds zero true positives and is not able to predict any failures that occurred. This is clearly shown in the resulting confusion matrixes in Figure 7.

On the other hand, by reweighting the SVM to weight the False Negative errors 10 times more than the False Positive errors, a non-trivial result (all data points assigned to the same class) was obtained. In addition, non-trivial results were obtained when applying the SMOTE-SVM approach. The F1 score for the weighted SVM is 0.533, while the F1 score for the SVM after borderline SMOTE-SVM is 0.444. Therefore, for this classification task, the weighted SVM achieves the highest overall performance in terms of the F1 score. The resulting confusion matrixes for the weighted SVM and SMOTE-SVM are given in Figure 7.

To interpret the confusion matrices, the different cells in the confusion matrix are labeled in Table 5, where negatives are non-failure, and positives are failures.

*Table 5: Labels of the confusion matrix*

| True Negatives | False positives |
|----------------|-----------------|
| False Negatives | True positives |

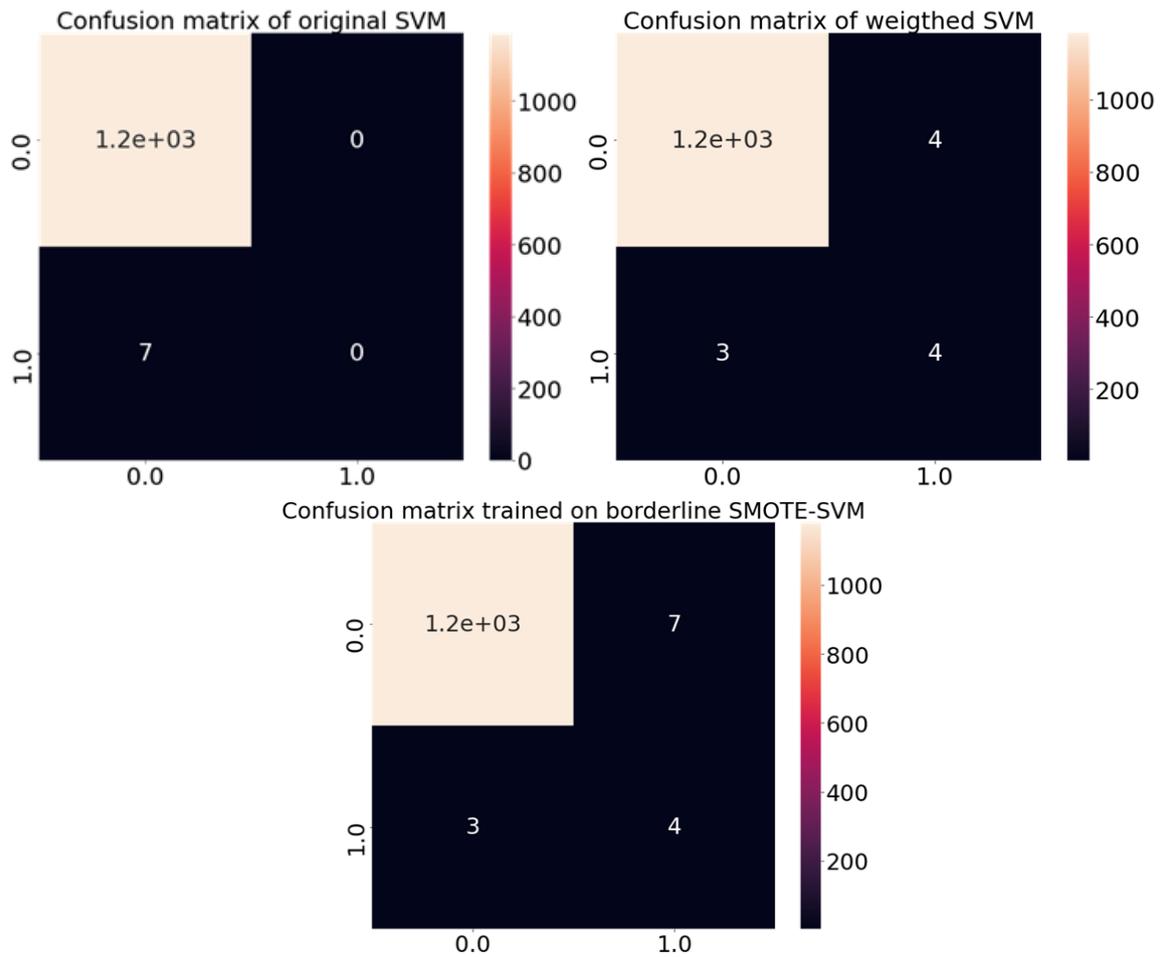

*Figure 7: Confusion matrixes for original SVM (upper left), weighted SVM on hourly data (upper right), and SVM trained on a synthetic dataset with borderline SMOTE-SVM.*

The resulting confusion matrices in Figure 7 shows that the original SVM (upper left) misses every failure. Hence, the original SVM achieves a sensitivity score of 0/7=0. Therefore, the original SVM is not the proper model for such a task.

The result for the weighted SVM (upper right in Figure 7) indicates an acceptable score in terms of sensitivity. Of 7 failures in the test set, the weighted SVM detect 4. This results in a sensitivity score of 4/7 = 0.57 or 57%. The weighted SVM only predicts 4 false positives.

The SVM trained with SMOTE, which synthetically augments the minority class, achieves the same sensitivity score as for the weighted SVM. However, the number of false positives is increased to 7. This shows that for this classification task, the SMOTE-SVM approach is not required to increase the classification performance and it is sufficient to weight the two classes differently in this case.

Given the complexity of the problem with the high-dimensional dataset, the large imbalance between classes, and the challenging case of distinguishing between failures and non-failures; a sensitivity score of 57% (i.e., a 57% accuracy on the failure class) and an F1 score of 0.533 is acceptable. In addition, the promising classification results in terms of F1-and sensitivity score show that the three EAs selected provide valuable information on how to narrow down the potential locations of the weather-related causes of failure, which has been a major challenge for the DSOs. Therefore, these findings have set a promising ground for a follow-up study on localizing the weather-related causing factors more in detail.

To detect the possible causes for the 7 failures in the test set, a closer investigation on the specific failures was performed in collaboration with the DSO. By investigating each failure individually, it was found that wind speed seems to be an important factor as there were several days with failures where the magnitude of wind was high (above 25 meters per second). This is in line with the pre-analysis from the distribution plots and the KL-divergence in section 4.2, which shows that wind speed could be an important factor.

For one of the 7 failures, there were not identified any specific causes. However, the official FASIT report for the specific incident shows that this failure was reported to be an *earth-failure.* In the case of an earth-failure, the electrical transformers connected to the grid break, and the power that flows through the transformer would flow to the ground (Earth). When the end of the electrical transformer station that contacts the ground level is on the downstream side, an earth-failure results [22].

Interestingly, the earth failure is one of the 3 false negatives. However, since it is independent of the weather and electricity load measure considered as input variables, it is correct that the SVM assigns it to the non-failure class and should not be considered as a true error.

In the end, out of 7 failures in the test set, 4 failures were correctly detected with the ML model. Among the 3 false negatives, one failure was identified as earth failure, and for the 2 remaining failures wrongly classified, the cause remains unknown.

As the individual investigations of the 7 failures showed that the magnitude of wind speed seems to be a causing variable, there would be of interest to categorize the different variables in terms of importance. This will provide knowledge regarding how the DSO could act in advance based on variables that are given as important, and which variables are interpreted as less important. Therefore, in the following section, the variables of importance for causing failures are identified by statistical tests and the weighted SVM.

## 5.2 Identification of important and non-important features

To gain knowledge of which variables are the most relevant for causing failures in the power grid, it is possible to rank the different variables in terms of how much they contribute to explaining the classification results obtained by the ML models. The significance of each feature is computed by performing statistical tests (F-test), where the p-value for each feature is reported with the corresponding weight of the weighted SVM [31]. In this study, the rank of the features is computed to detect the important variables. Figure 8 shows the score for each feature, where the features with the largest weight are the most important ones.

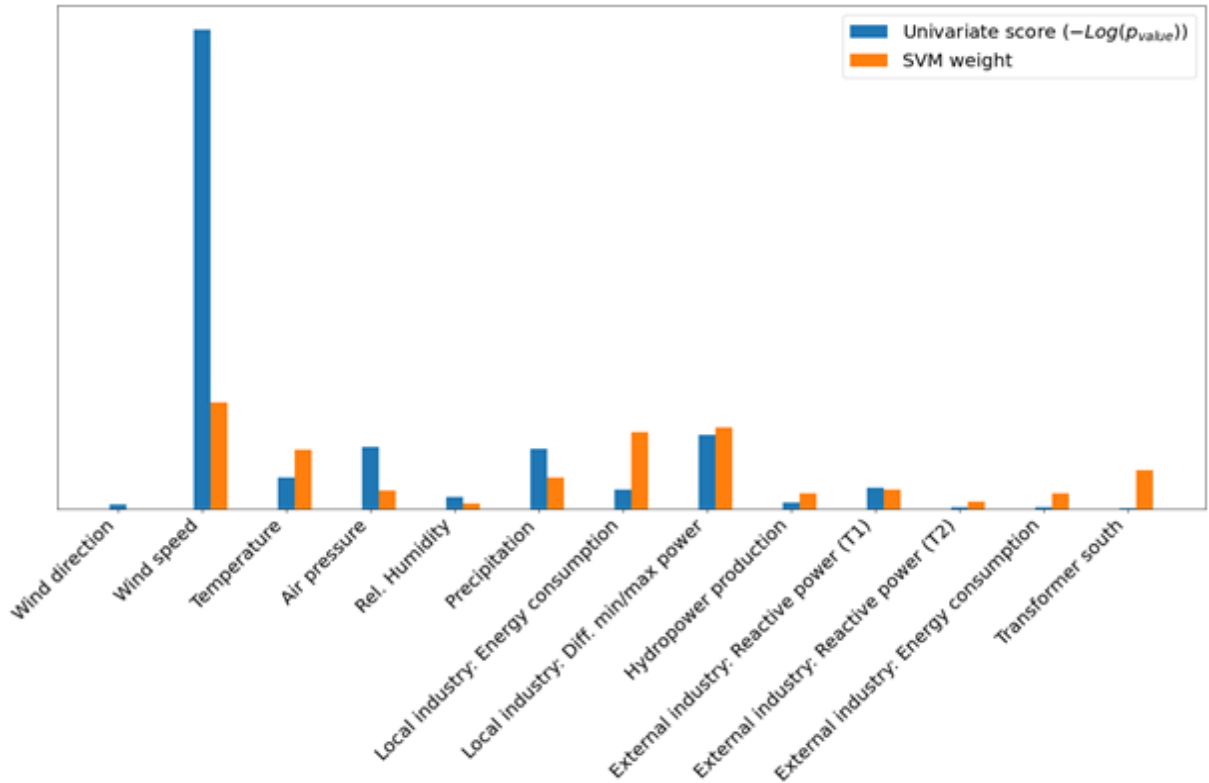

*Figure 8: Univariate feature selection of failure data. The blue bar represents the univariate score reported as the p-value. The orange bar represents the SVM weight for each feature*

From Figure 8, the rank of wind speed is by far the highest, indicating that wind speed is the most important variable in discriminating between the failure and non-failure class. This confirms the pre-analysis from the Gaussian KDE distributions and KL-divergences in section 4.2. This is also in line with experiences from the DSO and local knowledge, which states that they often experience failures when the magnitude of wind is large.

The second most important variable is the difference in minimum and maximum power in the local industry. This is in line with experiences from the local industries: In periods when the industry has high activity, there is an increased likelihood for failures. The third and fourth most important variables are air pressure and precipitation. Precipitation could be due to snow accumulation on the power lines or icing in the winter season. However, the impact of air pressure and precipitation needs dedicated investigation, as there is no clear direct link between the variables and failures. As a suggestion, closer investigations could be performed by the DSO that could install measurement instruments on the different components of relevance in the power grid.

The remaining non-weather-related variables (rightmost variables in Figure 8) contribute much less in explaining the classification result.

# 6 Conclusion

The contributions of our study can be summarized as follows. We have:

- Identified the main challenges in a specific real-world power grid in collaboration with the local DSO.

- Collected relevant meteorological historical data in the area of the energy grid.

- Collected energy (consumption-and production) data on the grid, both in periods of normal operating conditions and when the failures are recorded.

- Formulated a methodology to detect the failures based on the dataset that is constructed.

- Analysed the results in collaboration with the DSO to interpret our findings and evaluate the effectiveness of the proposed methodology.

We have addressed the challenges of detecting short-term failures in the power grid, which have major consequences for the local industries that are experiencing an increased frequency of failures in the grid. In total 13 different variables were collected (6 weather-related, and 7 energy-related). We assumed that the utility poles in the areas at high altitudes were more exposed to harsh weather conditions and that failures in the network due to weather are more likely to occur in these areas. In addition, we have considered that the large distances between the utility poles in periods with high wind could cause the lines to collide and lead to power interruptions. We focus on three such regions (EA1, EA2, and EA3) based on 1) altitude and 2) distance between utility poles. To find the altitude of the utility poles, a high-resolution altitude map of the case area was made, and the coordinates of the utility poles were mapped into the altitude map. The geodesic distance between the Lat-Long coordinates of the utility poles was calculated in order to find the distance between the utility poles. Relevant consumption and production variables were collected in collaboration with the DSO. To get a broader perspective, time series of energy consumption were extracted from an industry that is not directly connected to SVAN22LY1. This industry is the largest customer in the power grid, and it was thought that the consumption pattern here could affect the power quality of SVAN22LY1.

After the pre-analyses, different ML techniques were examined to classify and identify failures. However, the large imbalance in the dataset (7,958 normal conditions, and 44 failures), caused difficulties. 4 different one-class classification models were used. These resulted in a low F1 Score and were not further assessed. We obtained acceptable results with an SVM. The ordinary SVM model, on the other hand, failed to successfully classify our dataset. By weighting the SVM, we were able to classify the dataset with acceptable results. The weighted SVM model resulted in an F1 score of 0.533 and achieved a sensitivity result of 57% (identified 57% of the failures in the test set).

In terms of the F1 and the sensitivity scores, the results are promising since these show that the three EAs selected provide valuable information on how to narrow down the potential locations of weather-related causes of failure. These findings have set a promising ground for a follow-up study on localizing the weather-related contributing factors in more detail.

The potential impact on the power quality was explored by investigating high-resolution data of 1-minute from the external industry. The weighted SVM could not discriminate the classes at first due to high imbalance. By selecting 50,000 samples randomly from this dataset, we obtained a sensitivity score of 50%. SMOTE-SVM resulted in a low score for this dataset (see Supplementary chapter upon request by the authors for a full analysis).

The variables were ranked according to importance in terms of being a triggering factor for failures in the power grid. We have confirmed (KDE plot and KL divergence) that wind speed is the most important feature. Furthermore, we have identified the activity of local industries as another important parameter. The effects of air pressure and precipitation are something that needs to be investigated further. These results, in collaboration with the DSO, lay a foundation for investigating the possible impact of other variables when higher-resolution data is available.

## Acknowledgements

O.F.E and M.C acknowledge the support from the research project "*Transformation to a Renewable & Smart Rural Power System Community (RENEW)",* connected to the Arctic Centre for Sustainable Energy (ARC) at UiT-the Arctic University of Norway through Grant No. 310026.

We thank the Arva Power Company for providing valuable insight into the problem, and for providing the necessary datasets. We would especially like to thank Inga Setså Holmstrand and Sigurd Bakkejord at Arva Power Company for the valuable collaboration and discussions during the study.